\documentclass[a4paper,11pt]{article}
\usepackage{pos}
\newcommand{\dhd}{{\textstyle d} \lower.03ex\hbox{\kern-0.38em$^{\scriptstyle-}$}\kern-0.05em{}}

\title{TMD factorization bridging large and small x}

\author[a]{Swagato Mukherjee}
\author[b]{Vladimir Skokov}
\author[b,c]{Andrey Tarasov}
\author[b]{Shaswat Tiwari}

\affiliation[a]{Physics Department, Brookhaven National Laboratory\\
  Upton, New York 11973, USA}

\affiliation[b]{Department of Physics, North Carolina State University,\\
Raleigh, NC 27695, USA}

\affiliation[c]{Joint BNL-SBU Center for Frontiers in Nuclear Science (CFNS) at Stony Brook University, Stony Brook, New York 11794, USA}

\emailAdd{sstiwari@ncsu.edu}

\abstract{QCD factorization takes different forms in the large-x and small-x regimes. At large-x,  collinear factorization leads to the DGLAP evolution equation, while at small-x, rapidity factorization results in  the BFKL equation. To unify these different regimes, a new TMD factorization based on the background field method is proposed. This factorization not only reduces to CSS and DGLAP in the large-x limit and BFKL in the small-x limit, but also defines a general evolution away from these regimes.}

\FullConference{31st International Workshop on Deep Inelastic Scattering (DIS2024)\\
 8–12 April 2024\\
Grenoble, France\\}

\begin{document}
\maketitle

\section{Introduction}
In these proceedings, we shall summarize the results of \cite{Mukherjee:2023snp}.
In the TMD factorization approach \cite{Collins:1981uk,Collins:1984kg,Collins:1987pm,Collins:1989gx}, the production of a color-singlet state (Drell-Yan pair, Higgs and $W$ boson productions etc.) in an unpolarized hadron-hadron scattering reads
\begin{eqnarray}
&&\frac{d\sigma}{dQdyd^2q_\perp} = \sum_{ij} H_{ij}(Q, \mu) \int d^2b_\perp e^{iq_\perp b_\perp} f_i(x_a, b_\perp, \mu, \zeta_a)f_j(x_b, b_\perp, \mu, \zeta_b) + O\left(\frac{q^2_\perp}{Q^2}\right)
\,
\label{int:TMDff}
\end{eqnarray}
where $Q$ is an invariant mass of the final state, $y$ is its rapidity, and $q_\perp$ is the measured transverse momentum\footnote{We use the following notation $q_\perp b_\perp \equiv q_m b_m$.} and  $q_\perp^2 << Q^2$.
The sum goes over partons participating in the hard scattering which is defined by a hard function $H_{ij}$. The transverse-momentum dependent parton distribution functions (TMDPDFs) depend on an impact parameter variable $b_\perp$. The $x_a$ and $x_b$ are the longitudinal-momentum fractions of the colliding hadrons carried by the partons.

The functions in Eq. (\ref{int:TMDff}) depend on unphysical factorization scales, $\zeta$ and $\mu$, that separate the dynamical modes in the TMD factorization scheme. Dependence on these parameters can be studied by perturbative methods and is governed by the anomalous dimensions $\gamma^i_\mu$ and $\gamma^i_\zeta$ which form a part of the so called Collins-Soper equation \cite{Collins:1981uk,Collins:1981va}. One can then construct the solutions to these equations, and evolve the TMDPDF from initial scales, $(\mu_0, \zeta_0)$, to final scales, $(\mu, \zeta)$, using
\begin{eqnarray}
&&f_i(x, b_\perp, \mu, \zeta) = \exp\Big(\int^\mu_{\mu_0}\frac{d\tilde{\mu}}{\tilde{\mu}}\gamma^i_\mu(\tilde{\mu}, \zeta_0)\Big)\exp\Big(\frac{1}{2}\gamma^i_\zeta(\mu, b_\perp)\ln\frac{\zeta}{\zeta_0}\Big)f_i(x, b_\perp, \mu_0, \zeta_0)
\,.
\label{tmd-evolved}
\end{eqnarray}
The above evolution resums large logarithms $\ln(\mu^2 b^2_\perp)$ and $\ln(\zeta b^2_\perp)$. These logarithms are of the ultraviolet (UV) origin, and the infrared (IR) structure of the TMD factorization is encoded in the boundary term $f_i(x, b_\perp, \mu_0, \zeta_0)$.

The TMDPDF, $f_i(x, b_\perp, \mu_0, \zeta_0)$, is intrinsically nonperturbative. However, for small values of $b_\perp \ll \Lambda^{-1}_{\rm QCD}$ the TMDPDF can be expanded around $b_\perp=0$ and matched
onto collinear parton distribution functions (PDFs) \cite{Collins:1984kg,Collins:1981uw,Kang:2012em,Braun:2009mi} and for small values of x it can be expanded around $x=0$ and matched onto dipole amplitudes \cite{Dominguez:2010xd,Dominguez:2011wm}. While both these methods are well developed, one needs to sum over the series in $b_{\perp}$ (higher twist terms) and the series in x (sub-eikonal terms) to get the full IR structure of TMDs. This does not seem feasible. 

Hence we propose a new approach, where we study the IR structure of TMDPDF's at next to leading order (NLO) in the background field method. We work in the dilute limit, in the background of 2 gluons. 

\section{Background Field method}
The concept of factorization can be elegantly introduced within the background field method. Starting with a matrix element of an arbitrary operator $\mathcal{O}(\hat{C})$ which depends on quark and gluon fields,\footnote{Here we understand $C$ as a general notation for the quark and gluon fields.} one can write it as a functional integral over those fields
\begin{eqnarray}
&&\langle P_1 | \mathcal{O}(\hat{C}) |P_2\rangle = \int \mathcal{D}C~ \Psi^\ast_{P_1}(\vec{C}(t_f)) \mathcal{O}(C) \Psi_{P_2}(\vec{C}(t_i)) e^{iS_{\rm QCD}(C)} \,
\label{int:mel-form1}
\end{eqnarray}
where $\Psi_{P_2}$ and $\Psi_{P_1}$ are initial and final state wave functions at $t_i\to-\infty$ and $t_f\to \infty$.

The main logic of the background field approach is to split the field into different modes (using some factorization scale $\sigma$),
\begin{eqnarray}
&&C_\mu \to A_\mu + B_\mu \,.
\label{int:Bj-spt}
\end{eqnarray}
We integrate over the ``quantum" modes A while keeping the ``background" B fixed. 
 One can think about the ``background" fields as fields associated with the target, and the ``quantum" fields as fields of the hard scattering. One can then integrate over the ``quantum" modes in the functional integral, to get
\begin{eqnarray}
&&\langle P_1 | \mathcal{O}(\hat{C}) |P_2\rangle = \sum_i H_i(\sigma)\otimes \langle P_1|\mathcal{V}_i(\sigma)|P_2\rangle \,
\label{int:fact}
\end{eqnarray}
where $H_i(\sigma)$ is obtained by integration over the ``quantum" modes and the matrix element depends only on the ``background" modes. This is the statement of factorization in the language of background fields. 

In order to study the matrix element, we further split the background modes $B = B^q + B^{bg}$ using a scale $\sigma'$ and then integrate over $B^q$ modes to get
\begin{eqnarray}
&&\langle P_1|\mathcal{V}_i( \sigma)|P_2\rangle = \sum_j C_{ij}(\sigma, \sigma')\otimes \langle P_1|\mathcal{V}_j(\sigma')|P_2\rangle \,
\label{mel:ptb}
\end{eqnarray}
where $C_{ij}(\sigma,\sigma')$ was obtained by integration over the dynamical modes $B^q$ between scales $\sigma$ and $\sigma'$. This is the statement of evolution in the backgound field language. Here $\sigma$ acts as the UV scale and $\sigma'$ acts as the IR scale. We shall now apply this technique to TMD factorization.
\section{TMD factorization}
In the background field approach, the hard function is separated from the TMD modes by the scale $\mu_{UV}^2$ and the are collinear and anti-collinear modes (for the two colliding hadrons in Drell-Yan) of the TMD matrix element are given by the beam function,
\begin{equation}
\begin{split}
&\mathcal{B}_{ij}(x_B, b_\perp) \\&
= \int^\infty_{-\infty} dz^- e^{-i x_BP^+ z^-} \langle P,S|\bar{T}\{F^m_{-i}(z^-, b_\perp) [z^-, \infty]^{ma}_b\} T\{[\infty, 0^-]^{an}_0 F^n_{-j}(0^-, 0_\perp)\}|P,S\rangle\,.
\label{gop-def}
\end{split}
\end{equation}
In Eq. (\ref{gop-def}), $i$ and $j$ are Lorentz  indices. The adjoint Wilson lines are along the light-cone direction
\begin{eqnarray}
&&[x^-, y^-]_{z_\perp} = \mathcal{P}\exp\Big[ig \int^{x^-}_{y^-} dz^- A_-(z^-, z_\perp)\Big]\,.
\end{eqnarray}
The collinear and anti-collinear modes are separated by their rapidity, $y\:=\:\ln{k^+ \over k^-}$. Depending on how the modes are separated, there might be an intersection region that is populated by the so-called soft modes, which have the matrix element
\begin{eqnarray}
&&\mathcal{S}(b_\perp) = \frac{1}{N^2_c - 1}\langle 0|{\rm Tr}[S^\dag_{\bar{n}}(b_\perp)S_{n}(b_\perp) S^\dag_{n}(0_\perp) S_{\bar{n}}(0_\perp)]|0\rangle \,
\label{soft_func}
\end{eqnarray}
where the Wilson line
\begin{eqnarray}
&&S_{n}(b_\perp) = P\exp\Big[ig\int^\infty_0 d(x\cdot \bar{n}) n\cdot A(x\cdot \bar{n}, b_\perp)\Big]\, .
\end{eqnarray}
The full matrix element of the background modes is defined as
\begin{eqnarray}
&&f_{ij}(x_B, b_\perp) = \sqrt{\mathcal{S}(b_\perp)} \mathcal{B}_{ij}(x_B, b_\perp)\,.
\label{fTMD}
\end{eqnarray}
\section{NLO calculation}
\begin{figure}[htb]
 \begin{center}
\includegraphics[width=100mm]{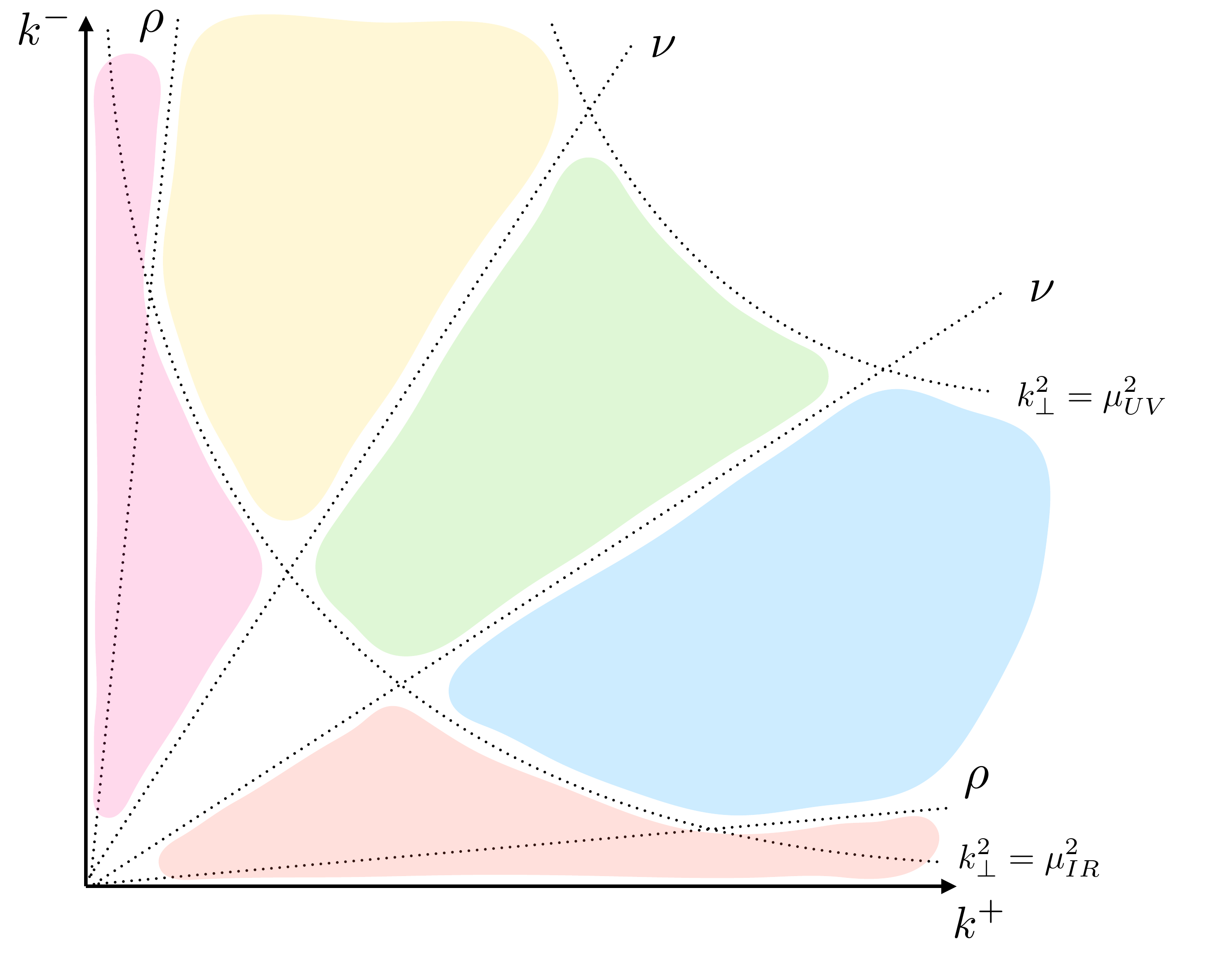}
 \end{center}
\caption{\label{fig:fnew}Illustration of our factorization scheme. Green is soft, blue is collinear. $\mu_{IR}$ and $\rho$ are IR cut-offs. $\nu$ defines the region of intersection between two collinear modes. The double counting of this region is removed by the inclusion of the soft factor $\mathcal{S}(b_\perp)$ in the factorization formula. A mass-shell hyperbolas are defined by a scale $\mu^2$ via $2k^+ k^- = k^2_\perp = \mu^2$.}
 \end{figure}
We calculate the one loop real and virtual corrections to the beam function and the soft function using the background field approach. The UV and IR divergences are regulated using dimensional regularization. This gives us the scales $\mu_{UV}$ and $\mu_{IR}$. The rapidity divergences are regulated using the $\eta$-regularization, which for the beam function takes the form
\begin{eqnarray}
&&\int^\infty_0 \frac{d k^-}{k^-} \to \nu^\eta \int^\infty_0 \frac{d k^-}{k^-}  |k^+ |^{-\eta} \,,
\label{eta-reg}
\end{eqnarray}
with $k^+\:=\:{k_{\perp}^2 \over 2 k^-}$ and for the soft function, 
\begin{eqnarray}
\int \frac{dk^+dk^-}{k^+k^-} = \nu^\eta 2^{-\eta/2} \int \frac{dk^+dk^-}{k^+k^-} |k^+ - k^-|^{-\eta} \,.
\end{eqnarray}
One encounters transverse UV and IR divergences in both real and virtual corrections. However the real corrections only diverge in the UV of rapidity ($k^- \rightarrow \infty$) while the virtual corrections diverge in the IR of rapidity ($k^- \rightarrow 0$).
Regularization of these rapidity divergences introduces scales $\nu$ for the rapidity UV divergence and the scale $\rho$ for rapidity IR divergence. The soft factor has a rapidity UV divergence and hence depends on the scale $\nu$ (see figure \ref{fig:fnew} for scale structure). The UV poles are then removed by multiplying the UV renormalization factor $Z_{UV}$ and the 1-loop soft factor $\mathcal{S}^{(1)}(b_{\perp})$ to the beam function $\mathcal{B}_{ij}(x,b_{\perp})$. The IR poles are absorbed into the nonpertubative $f_{ij}$ constructed from the background fields. Carrying out this procedure, one then gets the final result
\begin{eqnarray}
&&f_{ij}(x_B, b_\perp, \mu^2_{\rm UV}, \zeta) = f_{ij}(x_B, b_\perp, \mu^2_{\rm IR}, \rho) - 4\alpha_s N_c \int \dhd^2p_\perp e^{ip_\perp b_\perp} \int^1_0 \frac{dz }{z(1-z)} \int \dhd^2k_\perp 
\nonumber\\
&&\Big[ \mathcal{R}^{a}_{ij;lm}(z, p_\perp, k_\perp) + \mathcal{R}^{b}_{ij;lm}(z, p_\perp, k_\perp) \Big] \int d^2z_\perp e^{-i(p_\perp - k_\perp)z_\perp}f_{lm}(\frac{x_B}{z}, z_\perp, \mu^2_{\rm IR}, \rho) 
\nonumber\\
&&+ \frac{ \alpha_s N_c }{2\pi} \Big( - \frac{1}{2}(L^{\mu_{\rm UV}}_b)^2 + L^{\mu_{\rm UV}}_b \ln \frac{\mu^2_{\rm UV}}{\zeta^2}   - \frac{\pi ^2 }{12}\Big) f_{ij}(x_B, b_\perp, \mu^2_{\rm IR}, \rho) - \frac{\alpha_s N_c}{\pi } L^{\mu_{\rm IR}}_b 
\nonumber\\
&&\times \int^1_0 dz \Big[ \frac{1 }{(1-z)_+} + \frac{1 }{z} \Big] f_{ij}(\frac{x_B}{z}, b_\perp, \mu^2_{\rm IR}, \rho)  - \frac{\alpha_sN_c}{2\pi } \int d^2z_\perp \int \dhd^{2} p_\perp e^{ip_\perp (b-z)_\perp}
\nonumber\\
&&\times \Big( \frac{1}{2} \ln^2 \frac{\mu^2_{\rm IR}}{p^2_\perp } + \ln \frac{\mu^2_{\rm IR}}{p^2_\perp } 
\ln \frac{\rho}{\zeta}
- \frac{\pi^2 }{12} \Big) \frac{g_{il} p_j p_m + p_i p_l g_{mj}}{ p^2_\perp } f_{lm}(x_B, z_\perp, \mu^2_{\rm IR}, \rho)+\frac{\alpha_s N_c}{2\pi} \int d^2z_\perp  
\nonumber\\
&&\times \int \dhd^2p_\perp e^{ip_\perp (b - z )_\perp} \Big( \frac{\beta_0}{2N_c} \ln \frac{\mu^2_{\rm UV}}{p^2_\perp} + \frac{67}{18} - \frac{5N_f}{9N_c} \Big) f_{ij}(x_B, z_\perp, \mu^2_{\rm IR}, \rho) + O(\alpha^2_s)\,
\label{fr:form1}
\end{eqnarray}
with $L^\mu_b \equiv \ln\Big(\frac{b^2_\perp \mu^2}{4e^{-2\gamma_E}}\Big)$, $\mu$ is the $\overline{\rm{MS}}$ scale and $\zeta\:=\:x_B P^+$. 
It can be further shown that the equation $\label{fr:form1}$ reduces to the well-known collinear matching formula in the limit of small $b_{\perp}$
\begin{eqnarray}
&&f_1(x_B, b_\perp, \mu^2_{\rm UV}, \zeta) = f_1(x_B, 0_\perp, \mu^2_{\rm IR})- \frac{\alpha_s N_c}{\pi}  L^{\mu_{\rm IR}}_b \int^1_0 \frac{dz }{z} P_{gg}(z) f_1(\frac{x_B}{z}, 0_\perp, \mu^2_{\rm IR})
\label{cm:form4}\\
&&  + \frac{\alpha_s N_c}{2\pi } \Big( - \frac{1}{2}(L^{\mu_{\rm UV}}_b)^2 + L^{\mu_{\rm UV}}_b \ln \frac{\mu^2_{\rm UV}}{\zeta^2}   - \frac{\pi ^2 }{12}\Big) f_1(x_B, 0_\perp, \mu^2_{\rm IR})+\dots\,
\nonumber
\end{eqnarray}
where $f_1(x_B,b_{\perp})$ is the unpolarized TMDPDF and $P_{gg}(z)$ is the DGLAP splitting function. Similarly, in the small-x limit it can be shown that the equation $\label{fr:form1}$ reduces to
\begin{eqnarray}
&& f_1(x_B, p_\perp, \mu^2_{\rm UV}, \zeta) \simeq \mathcal{H}_1(p_\perp, \rho) + 
\ln \frac{\rho}{\zeta} 
\int \dhd^2 k_\perp K_{\rm BFKL}(p_\perp, k_\perp) \mathcal{H}_1(p_\perp - k_\perp, \rho)
\nonumber\\
&& + \frac{ \alpha_s N_c }{2\pi} \int d^2b_\perp \Big( - \frac{1}{2}(L^{\mu_{\rm UV}}_b)^2 + L^{\mu_{\rm UV}}_b \ln \frac{\mu^2_{\rm UV}}{\zeta^2}   - \frac{\pi ^2 }{12}\Big) \int \dhd^2 k_\perp e^{i k_\perp b_\perp } \mathcal{H}_1(p_\perp - k_\perp, \rho)
\nonumber\\
&&+ \frac{\alpha_s N_c}{2\pi} \Big( \frac{\beta_0}{2N_c} \ln \frac{\mu^2_{\rm UV}}{p^2_\perp} + \frac{67}{18} - \frac{5N_f}{9N_c} \Big) \mathcal{H}_1(p_\perp, \rho) \,
\label{sx:form4}
\end{eqnarray}
where $\mathcal{H}$ is the dipole amplitude and $K_{BFKL}$ is the well-known BFKL kernel.
\section{Conclusions}
In this paper, we study the IR structure of the gluon TMDPDFs. We go beyond the standard collinear matching procedure, which is performed in the $b_\perp \ll \Lambda^{-1}_{\rm QCD}$ approximation, and develop a new approach that is valid in a wide region of $x_B$ and $b_\perp \lesssim \Lambda^{-1}_{\rm QCD}$. To do that, we employ the background field method and perform the calculation of the TMDPDFs in the dilute limit at the NLO order. This gives us a general matching kernel, which reduces to the known collinear matching kernel and eikonal expansion results in the appropriate limits.

\section{Acknowledgnents}
We thank Ian Balitsky, Paul Caucal, Florian Cougoulic, Yuri V. Kovchegov, Farid Salazar, Bj\"orn Schenke, Tomasz Stebel, Yossathorn Tawabutr, and Raju Venugopalan for the discussions.\par
This material is based upon work supported by The U.S. Department of Energy, Office of Science, Office of Nuclear Physics through Contract Nos.~DE-SC0012704 and DE-SC0020081, and within the frameworks of Saturated Glue (SURGE) Topical Collaboration in Nuclear Theory. V.S. thanks the Binational Science Foundation grant \#2021789 for support.

\bibliographystyle{unsrt}
\bibliography{main}

\end{document}